\documentclass[12pt]{article}
\usepackage{amsmath}
\usepackage{amssymb}
\begin{document}
\vskip 2cm
\begin{center}
{\bf {\Large {Universal Superspace Unitary Operator for Some Interesting Abelian Models: Superfield Approach}}}

\vskip 3.0cm

{\sf T. Bhanja$^{(a)}$, N. Srinivas$^{(a)}$, R. P. Malik$^{(a,b)}$}\\
$^{(a)}$ {\it Physics Department, Centre of Advanced Studies,}\\
{\it Banaras Hindu University, Varanasi - 221 005, (U.P.), India}\\

\vskip 0.1cm

\vskip 0.1cm

$^{(b)}$ {\it DST Centre for Interdisciplinary Mathematical Sciences,}\\
{\it Faculty of Science, Banaras Hindu University, Varanasi - 221 005, India}\\
{\small {\sf {e-mails: tapobroto.bhanja@gmail.com; seenunamani@gmail.com; rpmalik1995@gmail.com}}}

\end{center}

\vskip 2cm
\noindent
{\bf Abstract:} Within the framework of augmented version of superfield formalism, 
we derive the superspace unitary operator and show its usefulness in the derivation
 of Becchi-Rouet-Stora-Tyutin (BRST) and anti-BRST symmetry transformations for a
 set of interesting models for the Abelian 1-form gauge theories. These models are 
(i) a one (0+1)-dimensional (1$D$) toy model of a rigid rotor, (ii) the two 
(1+1)-dimensional (2$D$) modified versions of the Proca and anomalous Abelian 1-form 
gauge theories, and (iii) the 2$D$ self-dual bosonic gauge field theory. We provide, in some
 sense, the alternatives to the horizontality condition (HC) and the gauge invariant
 restrictions (GIRs) in the language of the above superspace (SUSP) unitary operator. 
One of the key observations of our present endeavor is the result that the SUSP unitary
operator and its hermitian conjugate are found to be the {\it same} for {\it all} 
the Abelian models under consideration (including the 4$D$ {\it interacting}  Abelian 
1-form gauge theories with Dirac and complex scalar fields which have been discussed earlier).
Thus, we establish the {\it universality} of the SUSP operator for the above Abelian theories.

\vskip 2cm

\noindent
PACS numbers: 11.30.Pb; 03.65.-w; 11.30.-j\\

\noindent
{\it {Keywords}}: {Abelian models, augmented superfield/supervariable formalism, (anti-)BRST symmetries,
superspace unitarity operator, horizontality condition (HC), gauge invariant restrictions (GIRs),
alternative to HC and GIRs, universality of the SUSP operator}

\newpage
\section{Introduction}

The usual superfield approach [1-8] to Becchi-Rouet-Stora-Tyutin (BRST) formalism is a 
geometrically rich and physically very intuitive method which sheds light on the 
{\it physical} interpretation of the nilpotency and absolute anticommutativity 
property of the proper (anti-)BRST symmetries.
These properties are sacrosanct because the nilpotency property encodes the
fermionic nature of the (anti-)BRST symmetries and the absolute anticommutativity property
represents the linear independence of the BRST and anti-BRST symmetries. In addition,
the above superfield approach also leads 
to the derivation of the precise form of the (anti-)BRST symmetries and associated Curci-Ferrari (CF) condition in the 
description of the $p$-form ($p$ = 1, 2, 3, ...) (non-)Abelian gauge theories (see, e.g., [4-6,9,10] for details).
However, this approach is useful {\it only} in the derivation of the (anti-)BRST symmetries for the $p$-form gauge 
and associated (anti-)ghost fields of a given $p$-form gauge theory. It does not shed any light on the 
derivation of the (anti-)BRST symmetries that are associated with the {\it matter} fields in 
a given {\it interacting} $p$-form gauge theory.

The above superfield approach has been consistently  generalized so as to derive the (anti-)BRST 
symmetries for the gauge, associated (anti-)ghost and {\it matter} fields of a given {\it interacting}
(non-)Abelian 1-form gauge theory by invoking the horizontality condition (HC) and the gauge 
(i.e. (anti-)BRST) invariant restrictions (GIRs) on the superfields defined on the 
($D$, 2)-dimensional supermanifold corresponding to a given $D$-dimensional interacting
(non-)Abelian 1-form gauge theory defined on the $D$-dimensional flat Minkowski spacetime 
ordinary manifold (see, e.g., [11-14]). The consistently generalized version of the superfield
approach to BRST formalism [4-6] has been christened as the augmented version of superfield approach to 
BRST formalism in our earlier works [11-14].

In the seminal works [4-6], a superspace (SUSP) unitary operator has been introduced to derive 
the (anti-)BRST symmetries associated with the gauge field, (anti-)ghost fields and a {\it generic} matter field 
in the context of the superfield approach to BRST formalism. This SUSP unitary operator has been 
introduced so as to maintain the explicit group structure in the SUSP transformation space of the  fields of a 
given 4$D$ non-Abelian 1-form  gauge theory. However, the precise expressions  for this 
SUSP unitary operator and its hermitian conjugate have {\it not} been explicitly derived 
mathematically. Rather, these expressions have been chosen intelligently. The hermitian
conjugate operator (corresponding to a SUSP unitary operator) has been derived after imposing
some hermitian conjugation  conditions from {\it outside} on the fields as well as  the 
Grassmannian parameters of the superfield version of the theory. The above SUSP operators, 
besides maintaining the group structure, provide the alternatives to the HC and GIRs 
as we shall see in our further discussions.

In a very recent couple of papers [15,16], we have been able to derive explicitly the expressions for 
the SUSP unitary operator and its hermitian conjugate {\it without} imposing any outside  hermitian
conjugation conditions on the fields and/or the Grassmannian variables of the supersymmetrized version 
of the {\it interacting} 4$D$ Abelian and (non-)Abelian 1-form gauge theories. 
In the former case, we have discussed the QED 
with Dirac and complex scalar fields coupled to the Abelian 1-form gauge field [15] and, 
in the latter case, we have considered the 4$D$ non-Abelian interacting theory with Dirac fields. 
The purpose of our present  endeavor is to show the {\it universality} of the above SUSP unitary
operator and its hermitian conjugate in the description of the 1$D$ and 2$D$ Abelian 1-form gauge theories 
of different kinds where the covariant derivatives are {\it not} explicitly defined. 
For this purpose, we take into consideration  the 1$D$ toy model of a rigid rotor, the 2$D$ modified
versions of the Proca as well as the anomalous gauge theories and the 2$D$ self-dual bosonic theory. 
These models are very {\it interesting} Abelian 1-form theories in one and two dimensions of
 spacetime because these have been proven to provide the physical 
examples of Hodge theory (see, e.g., [17-23]) where there exist many interesting {\it internal} 
symmetries (and corresponding conserved charges). To be precise, the models
under considerations respect the proper (anti-)BRST and (anti-)co-BRST symmetry
transformations, a unique bosonic symmetry transformation and the ghost-scale symmetry 
transformations which together provide the physical realizations of the cohomological 
operators of differential geometry.

In our present investigation, we have applied the augmented version of superfield/supervariable
formalism to derive the (anti-)BRST symmetries for the 2$D$ and 1$D$ Abelian 1-form gauge
theories and expressed these results in terms of the SUSP unitary operators. We have shown
that the mathematical form of the SUSP unitary operator and its hermitian conjugate is 
{\it universal} for the 1$D$, 2$D$ and 4$D$ 1-form gauge theories. In fact, we have obtained the
universal form of the  above operators which maintain  the $U(1)$ group structure in 
the SUSP transformation space of the gauge variables/fields  and {\it other} variables/fields
of these theories. Furthermore, the SUSP unitary operators have  also been shown to provide the alternatives 
to the HC and GIRs that enable us to derive the {\it full} set of proper (anti-)BRST symmetry
transformations for {\it all} the variables/fields of a given theory. The {\it universality} of the 
mathematical form of the SUSP unitary operator (and its hermitian conjugate) is one of the highlights
of our present investigation where we have considered  different kinds of Abelian 1-form gauge
theories.

Our present investigation is essential because it is motivated by the following key factors. 
First and foremost, we have demonstrated the {\it universality} of the SUSP unitarity
operator (and its hermitian conjugate) for {\it all} the Abelian 1-form gauge theories 
defined on the 1$D$, 2$D$ and 4$D$ flat background Minkowski spacetime manifolds. This 
observation is one of the decisive features of our present investigation. Second, the existence
of the SUSP unitarity operator (and its hermitian conjugate) provides the alternatives 
to the HC and the GIRs where the group structure is very explicitly and elegantly  maintained. 
This is a very nice feature of the SUSP unitary operator and its hermitian conjugate. 
This group structure is somewhat hidden when we exploit the 
potential and power of the HC and the GIRs. Third, the Abelian models under considerations
are {\it interesting} because they provide a set of tractable examples of the
Hodge theory within the framework of BRST formalism [17-23]. Finally, our present work is our modest
first step towards our central goal of providing a theoretical {\it generality} for
the existence of the SUSP unitarity operator and its hermitian conjugate for the interacting
non-Abelian gauge theories as well.

The material of our present investigation is organized as follows. 
In Sec. 2, to set up the convention and notations, we discuss briefly the 
(anti-)BRST symmetries in the Lagrangian formulation for the one (0 + 1)-dimensional 
(1$D$) toy model of a rigid rotor, 2$D$ modified versions of the Proca theory
 as well as the anomalous gauge theory and 2$D$ self-dual bosonic field theory:
 Our Sec. 3 is devoted to a concise description of the HC and the GIRs which enable us
 to derive the above (anti-)BRST symmetry transformations within the framework 
of augmented superfield formalism. Sec. 4 deals with the derivation of the results,
 obtained in Sec. 3, in terms of the superspace unitarity operator and its hermitian
 conjugate. Finally, we summarize our key results, make some concluding remarks and
 point out a few future directions for further investigations in Sec. 5

{\it General Notations and  Convention}: We denote the (anti-)BRST symmetries for 
{\it all} the Abelian models by the symbols 
$s_{(a)b}$ in our present endeavor. For the 2$D$ theories, we adopt the convention 
such that the 2$D$ background Minkowski spacetime manifold is endowed with the metric 
$\eta_{\mu\nu} $ with the signatures (+1, -1) so that the dot product between two 
non-null vectors $A_{\mu} $ and $B_{\mu} $ is denoted by $A \cdot B = A_{\mu} B^{\mu} 
= \eta_{\mu\nu} A^{\mu} B^{\nu} = A_{0} B_{0} - A_{i} B_{i}$ where the Greek indices
$\mu, \nu, \lambda, ....= 0, 1 $ stand for the spacetime directions and the Latin indices
$i, j, k, .....= 1 $ correspond to the space direction only. We choose the 2$D$  Levi-Civita 
tensor $\varepsilon_{\mu\nu}$ (with $ \varepsilon_{0 1}\,=\,+ 1\,=\,-\,\varepsilon^{0 1}$) which satisfy
 $ \varepsilon_{\mu\nu}\,\varepsilon^{\mu\nu}\,=\,2\,!, \, \varepsilon_{\mu\nu}\,
\varepsilon^{\nu\lambda}\,=\,\delta _{\mu}^{\lambda},\, 
\varepsilon_{\mu\nu}\,\varepsilon^{\mu\lambda}\,=\,-\,\delta _{\nu}^{\lambda}$.
We denote the scalar field by $\phi (x, t)$ in the cases of the 2$D$ modified versions of Proca theory,
anomalous gauge theory and the 2$D$ self-dual chiral bosonic theory. The corresponding superfield
has been denoted by $\Phi (x, \theta, \bar\theta)$ within the framework of  superfield formalism.

\section{Preliminaries: (Anti-)BRST Symmetries}

We discuss here, first of all, the (anti-)BRST symmetries $s_{(a)b} $ in the context of 
a 1$D$ toy model of a rigid rotor  (with mass $m = 1$) which is described by the following
first-order (anti-)BRST invariant  Lagrangian (see, e.g., [24,17-19] for details)
\begin{eqnarray}
L_b = \dot r \, p_r + \dot\vartheta \, p_{\vartheta} - \frac{p_{\vartheta}^2}{2 r^2} - 
\lambda \,(r - a) + B\,(\dot \lambda - p_r)
+ \frac{B^2}{2} - i\, \dot{\bar C}\,\dot C + i\, \bar C\, C,
\end{eqnarray}
where $(r, \vartheta)$ is a pair of polar coordinates, the pair $(p_{r}, p_{\vartheta})$ 
corresponds to the conjugate momenta with respect to $ (r, \vartheta)$,  $\lambda(t)$ 
is the Lagrange multiplier that turns out to be the analogue of the ``gauge'' variable, 
B(t) is the Nakanishi-Lautrup type auxiliary variable and the fermionic 
$(C^{2} = \bar{C}^{2} = 0, C \bar{C} + \bar{C} C = 0) $ (anti-)ghost variables 
$(\bar{C}) C $ are needed for the sake of unitarity in our theory. The above Lagrangian
(1) respects the following supersymmetric type off-shell nilpotent $(s_{(a)b}^{2} = 0)$
and absolutely anticommuting $(s_{b}s_{ab} + s_{ab}s_{b} = 0) $ (anti-)BRST symmetry
transformation $s_{(a)b} $:
\begin{eqnarray}
&& s_{b}\,(r, \vartheta, p_{\vartheta}, B, C)  = 0,\qquad s_{b}\, \lambda = 
\dot{C},\qquad  s_{b}\, p_{r} = - C,\qquad s_{b}\,\bar{C} = i B, \nonumber\\ 
&& s_{ab}\,(r, \vartheta, p_{\vartheta}, B, \bar{C})  = 0,\qquad s_{ab}\, \lambda =
 \dot {\bar{C}},\qquad  s_{ab}\, p_{r} = -\bar{C},\qquad  s_{ab}\, C = -i B.
\end{eqnarray}
As a consequence, the action integral $ S = \int dt\, L_{b}$  remains invariant under 
$s_{(a)b}$ (because  $s_b\, L_b = \frac{\partial}{\partial t} \, \big[B\,\dot C - (r - a)\,C\big], 
s_{ab}\, L_b = \frac{\partial}{\partial t} \, \big[B\,\dot {\bar C} - (r - a)\,\bar C\big]$). 
In our discussion, all the variables are function of the evolution parameter 
$t$ and the pair $(\dot{r} = {dr}/{dt}, \dot{\vartheta} = {d\vartheta}/{dt})$ stands for the
generalized radial and angular velocities.

We now focus on the Stuckelberg modified version of the 2$D$ Proca theory whose proper
(anti-)BRST invariant Lagrangian density is (see, e.g., [20,21] for details)
\begin{eqnarray}
\mathcal L_{B} &=& -\frac{1}{4}\,F_{\mu\nu}\, F^{\mu\nu} + \frac{m^{2}}{2}\, A_{\mu}\, A^{\mu} 
+ \frac{1}{2}\, \partial_{\mu} \phi \, \partial^{\mu}\phi  - m\, A_{\mu}\, \partial^{\mu} \phi \nonumber\\
 & + & B (\partial \cdot A + m\, \phi) + \frac {B^{2}} {2}
- i \,\partial_{\mu} \bar{C}\, \partial^{\mu} C + i\, m^{2}\, \bar{C}\, C,
\end{eqnarray}
where $F_{\mu\nu} = \partial_{\mu}\,A_{\nu}\,-\,\partial_{\nu}\,A_{\mu}$ is the curvature
 tensor that is derived from the curvature 2-form 
$F^{(2)}\,=\,d A^{(1)}\,\equiv \,[{(dx^{\mu}\,\wedge \,dx^{\nu})}/{2\,!}]\,F_{\mu\nu} $.
 Here $d\,=\,dx^{\mu}\,\partial_{\mu}$ is the exterior derivative (with $ d^{2} = 0 $)
 and the 1-form $ A^{(1)}\,=\,dx^{\mu}\,A_{\mu}$ defines the $U(1)$ vector potential 
$A_{\mu}$. The Stueckelberg field $ \phi $ (with mass $ m $) has been invoked to convert
the second-class constraints of the 2$D$ Proca theory (with mass $ m $) into the first-class
constraints. We have the (anti-)ghost fields $ (\bar{C})\, C $, too, in our BRST invariant
theory for the proof of unitarity. It is elementary to check that the following  off-shell 
nilpotent $(s_{(a)b}^{2} =\,0)$ and absolutely anticommuting $ (s_{b}\,s_{ab} + s_{ab}\,s_{b} = 0)$
(anti-)BRST symmetry transformations $ s_{(a)b}$
\begin{eqnarray}
&& s_{b}\,A_{\mu}\,=\,\partial_{\mu}\,C, \qquad \,\, s_{b}\,\phi\,  = \,m\,C, \qquad
 \,\, s_{b}\,\bar{C}\, = \,i\,B, \qquad \quad \,\, s_{b}\,(B, F_{\mu\nu}, C)\, = \,0,\nonumber\\
&& s_{ab}\,A_{\mu}\, = \,\partial_{\mu}\,\bar{C}, \qquad s_{ab}\,\phi\,  = \,m\,\bar{C}, 
\qquad s_{ab}\,C\, = \,-\,i\,B, \qquad s_{ab}\,(B, F_{\mu\nu}, \bar{C})\, = \,0,
\end{eqnarray}
leave the action integral $ S\,=\,\int d^{2}x\,\mathcal{L}_{B}$ invariant
for the physically well-defined fields which vanish off at $x\,=\pm \infty$
 because $s_b\, {\cal L}_B = \partial_\mu \, (B\, \partial^\mu \, C)$ and 
$s_{ab}\, {\cal L}_B = \partial_\mu \, (B\, \partial^\mu \, \bar C)$.

The next Abelian model of interest is the 2$D$ self-dual bosonic field theory whose precise
(anti-)BRST invariant Lagrangian density is (see, e.g., [22,23])
\begin{eqnarray}
\mathcal {L}_{B}^{(s)}\,&=&\,\frac{1}{2}\,\dot{\phi}^{2}\,-\,\frac{1}{2}\,\dot{v}^{2}\,
+\,\dot{v}\,(v'\,-\,\phi')\,+\,\lambda\,[\dot{\phi}\,-\,\dot{v}\,+\, v'\,-\,\phi']\nonumber\\
& - &\,\frac{1}{2}\,(\phi'\,-\,v')^{2} + \,B\,(\dot{\lambda}\,-\,v\,-\,\phi)\,+\,\frac{B^{2}}{2}\,
-\,i\,\dot{\bar{C}}\,\dot{C}\,+\,2\,i\,\bar{C}\,C,
\end{eqnarray}
where $ \dot{\phi}\,=\,\partial{\phi}/{\partial t}$, $\dot{v}\,=\,{\partial{v}}/{\partial t},
\dot{\lambda}\,=\,{\partial{\lambda}}/{\partial t}$, etc., are the generalized ``velocities" 
w.r.t the evolution parameter $ t$ of our present theory. We also follow the notation: 
$\phi'\,=\,{\partial{\phi}}/{\partial {x} } $, $v'\,=\,{\partial{v}}/{\partial x } $ which
is nothing but the single space derivative on the 2$D$ self-dual field $ \phi(x, t) $ and 
Wess-Zumino field $v(x, t)$. As explained earlier, $B (x, t)$ is the Nakanishi-Lautrup type 
auxiliary field and $ (\bar{C})\,C$ are the (anti-)ghost fields. It is straightforward 
to check that under the following  off-shell nilpotent and absolutely anticommuting 
(anti-)BRST symmetry transformations (see, e.g. [22,23] for details)
\begin{eqnarray}
&& s_{ab}\,\lambda\,=\,\dot{\bar{C}},\,\, s_{ab}\,\phi\,=\,-\,\bar{C},\quad
s_{ab}\,v\,=\,-\,\bar{C}, \quad s_{ab}\,C\,=\,-\,i\,B, \quad s_{ab}\,(B, \bar{C})\,=\,0, \nonumber\\
&& s_{b}\,\lambda \,=\,\dot{C},\quad s_{b}\,\phi \,=\,-\,C, \qquad
s_{b}\,v\,=\,-\,C, \qquad s_{b}\,\bar{C}\,=\,i\,B, \quad s_{b}\,(B, C)\,=\,0,
\end{eqnarray}
the Lagrangian density $ \mathcal{L}_{B}^{(s)}$ transforms as: 
$ s_{b}\,\mathcal{L}_{B}^{(s)}\,=\, \frac{\partial}{\partial t}[B\,\dot{C}] $, 
$ s_{ab}\,\mathcal{L}_{B}^{(s)}\,=\,\frac{\partial}{\partial t}[B\,\dot{\bar{C}}]$ 
thereby leaving the action integral $ S\,=\,\int d^{2}x\,\mathcal{L}_{B}^{(s)}$ 
invariant for the physically well-defined fields which vanish off at $ t\,=\,\pm \infty $.

Finally, we concentrate on the modified version of a 2$D$ anomalous Abelian 1-form gauge
theory (see, e.g., [25]) in its bosonized version. In this context, the (anti-)BRST
invariant Lagrangian density for this Abelian system (with electric charge $e\,=\,1$) is [25]
\begin{eqnarray}
\mathcal{L}_{B}^{(a)}\,&=&\,-\frac{1}{4}\,F^{\mu\nu}\,F_{\mu\nu}\,+\,\frac{1}{2}\,
\partial_{\mu}\phi\,\partial^{\mu}\phi\,+\,\frac{a}{2}\,A_{\mu}\,A^{\mu}\,+
\,(\eta^{\mu\nu}\,-\,\varepsilon^{\mu\nu})\,\partial_{\mu}\phi\,A_{\nu}\nonumber\\
& + &\,\sigma \,[(a\,-\,1)\,(\partial \cdot A)+ \,\varepsilon^{\mu\nu}\,\partial_{\mu}\,A_{\nu}]\,
+\,\frac{(a\,-\,1)}{2}\,\partial_{\mu}\sigma\,\partial^{\mu}\sigma\nonumber\\
& + &\,B\,(\partial\cdot A)\,+\,\frac{B^{2}}{2}\,-\,i\,\partial_{\mu}\bar{C}\,\partial^{\mu}C,
\end{eqnarray}
where the Abelian 1-form $A^{(1)}\,=\,dx^{\mu}\,A_{\mu}$ defines the vector potential
$ A_{\mu} $ and the curvature 2-form $ F^{(2)}\,=\,dA^{(1)} $
defines $ F_{\mu\nu}\,=\,\partial_{\mu}\,A_{\nu}\,-\,\partial_{\nu}\,A_{\mu}$ 
[which has only electric field (i.e. $F_{01} = E \equiv -\, \varepsilon^{\mu\nu}\, \partial_\mu \, A_\nu $)
as its existing component in 2$D$]. In the above, 
$ a $ is the parameter of ambiguity in the regularization of the fermionic determinant 
when the 2$D$ chiral Schwinger  model (with electric charge $e\,=\,1$) is bosonized 
in terms of the scalar field $ \phi$. A bosonic field $\sigma(x)$ has been introduced
 to convert the second-class constraints of the chiral Schwinger  model into the first-class 
constraints. As discussed earlier, the proper (anti-)BRST invariant Lagrangian density (7)
 contains the fermionic (anti-)ghost fields $ (\bar{C})\, C$ and the Nakanishi-Lautrup 
type auxiliary field $B(x)$. Under the following off-shell nilpotent $(s^{2}_{(a)b}\,=\,0)$
 and absolutely anticommuting $(s_{b}s_{ab}\,+\,s_{ab}s_{b}\,=\,0)$ (anti-)BRST symmetry transformations
\begin{eqnarray}
&& s_{ab}\,A_{\mu}\,=\,\partial_{\mu}\,\bar{C},\,\, s_{ab}\,C\,=\,-\,i\,B,\,\, s_{ab}\,\phi\,
=\,-\,\bar{C},\,\, s_{ab}\,\sigma\,=\,\bar{C},\,\, s_{ab}\,[B, \bar{C}, F_{\mu\nu}]\,=\,0,\nonumber\\
&& s_{b}\,A_{\mu}\,=\,\partial_{\mu}\,C, \quad s_{b}\,\bar{C}\,=\,i\,B, \quad s_{b}\,\phi\,
=\,-\,C,\quad s_{b}\,\sigma\,=\,C,\quad s_{b}\,[B, C, F_{\mu\nu}]\,=\,0,
\end{eqnarray}
the Lagrangian density (7) transforms as: 
\begin{eqnarray}
&& s_{ab}\,\mathcal{L}_{B}^{(a)}\,=\,\partial_{\mu}\,[B\,\partial^{\mu}\,\bar{C}\,+
\,(a\,-\,1)\,(\sigma \,\partial^{\mu}\,\bar{C}\,+ \,A^{\mu}\,\bar{C})\,-
\,\varepsilon^{\mu\nu}\,(\phi\,\partial_{\nu}\bar{C}\,-\,A_{\nu}\,\bar C)],\nonumber\\
&& s_{b}\,\mathcal{L}_{B}^{(a)}\,= \,\partial_{\mu}\,[B\,\partial^{\mu}\,{C}\,+
\,(a\,-\,1)\,(\sigma \,\partial^{\mu}\,{C}\,+\,A^{\mu}\,{C})\,
-\,\varepsilon^{\mu\nu}\,(\phi\,\partial_{\nu}\,C\,-\,A_{\nu}\, C)].
\end{eqnarray}
As a consequence, the action integral $S\,=\,\int d^{2}x\,\mathcal{L}_{B}^{(a)}$ 
remains invariant under the nilpotent (anti-)BRST symmetry transformations (8) for the physically
well-defined fields which vanish off at $x\,=\,\pm \infty$ when we apply the Gauss divergence theorem.

\section{(Anti-)BRST Symmetries: Superfield Approach}

In this section, we derive the proper (i.e. off-shell nilpotent and absolutely
anticommuting) (anti-)BRST symmetry transformations (for {\it all} the Abelian 
models under consideration) by exploiting the potential and power of HC and GIRs.
Towards this goal in mind, first of all, we generalize the basic variables 
$\lambda(t),\,C(t),\,\bar{C}(t)$ of the 1$D$ rigid rotor onto a (1, 2)-dimensional supermanifold,
parametrized by the superspace coordinate $Z^{M}\,=\,(t, \theta, \bar{\theta})$, as follows 
(see, e.g., [17-19] for details)  
\begin{eqnarray}
&&\lambda(t)\,\rightarrow \Lambda(t, \theta, \bar{\theta})\,=\,\lambda(t)\, +
 \,\theta\,\bar{R}(t)\,+\,\bar{\theta}\,R(t)\,+\,i\,\theta\,\bar{\theta}\, S(t), \nonumber\\
&& C(t)\,\rightarrow F(t, \theta, \bar{\theta})\,=\,C(t)\,+\,i\,\theta\,\bar{B}_{1}(t)\,+
\,i\,\bar{\theta}\,B_{1}(t)\,+\,i\,\theta\,\bar{\theta}\,s(t),\nonumber\\
&& \bar{C}(t)\,\rightarrow \bar{F}(t, \theta, \bar{\theta})\,=\,\bar{C}(t)\,+
\,i\,\theta\,\bar{B}_{2}(t)\,+\,i\,\bar{\theta}\,B_{2}(t)\,+\,i\,\theta\,\bar{\theta}\,\bar{s}(t),
\end{eqnarray}
where the secondary variables $(R,\bar{R}, s, \bar{s})$ and $(B_{1},\bar{B}_{1}, B_{2},\bar{B}_{2}, S)$
are fermionic and bosonic in nature because of the fermionic $(\theta^{2}\,=\,\bar{\theta}^{2}\,=
\,0 $, $\theta\, \bar{\theta}\,+\,\bar{\theta}\, \theta\,=\,0)$
nature of the pair of Grassmannian variables $(\theta,\,\bar{\theta})$ of the
superspace coordinate $Z^{M}\,=\,(t, \theta, \bar{\theta})$.
To determine the {\it exact} form of  the above secondary variables in terms 
of the basic and auxiliary variables of the Lagrangian (1) for the 1$D$ rigid 
rotor, we have to exploit the potential and power of the HC defined on the (1, 2)-dimensional supermanifold.

We define an Abelian  1-form $\lambda^{(1)}\,=\,dt\,\lambda(t)$ so that a null 2-form 
$d\lambda^{(1)}\,=\,0$ (where $d\,=\,dt\,\partial_{t}$ is the exterior derivative) could 
be constructed on a 1$D$ ordinary manifold. This can be generalized onto an appropriate 
(1, 2)-dimensional supermanifold as
\begin{eqnarray}
d\lambda^{(1)}\,\rightarrow\, \tilde{d}\tilde{\lambda}^{(1)}\,& = &\,[dt\,\partial_t\,
+\,d\theta\, \partial_{\theta}\,+\,d\bar{\theta}\,\partial_{\bar{\theta}}\,]\,\wedge\,
[\,dt\,\Lambda(t, \theta, \bar{\theta})\nonumber\\
& + &\,d\theta\,\bar{F}(t, \theta, \bar{\theta})\,+\,d\bar{\theta}\,F(t, \theta, \bar{\theta})\,],
\end{eqnarray}
where we have used the following generalizations
\begin{eqnarray}
&&d\,\rightarrow\,\tilde{d}\,=\,dt\,\partial_{t}\,+\,d\theta\,\partial_{\theta}\
,+\,d\bar{\theta}\,\partial_{\bar{\theta}},\nonumber\\
&&\lambda^{(1)}(t)\,\rightarrow\,\tilde{\lambda}^{(1)}(t, \theta, \bar{\theta})\,
=\,dt\,\Lambda(t, \theta, \bar{\theta})\,+\,d\theta\,\bar{F}(t, \theta, \bar{\theta})\,
+\,d\bar{\theta}\,F(t, \theta, \bar{\theta}),
\end{eqnarray}
in addition to the generalizations in (10). We demand the equality 
$\tilde{d}\tilde{\lambda}^{(1)}\, =\,d\lambda^{(1)}\,=\,0$ due to the requirement of the HC. 
This condition yields the following relationships between the secondary variables and the 
basic as well as the auxiliary variables of the Lagrangian (1) (see, e.g., [17-19] for details):
\begin{eqnarray}
&& R\,=\,\dot{C}, \qquad \bar{R}\,=\,\dot{\bar{C}},\qquad S\,=\,\dot{B},\qquad 
\qquad \,\,\bar{B}_{2}\,=\,0,\nonumber\\
&& B_{1}\,=\,0,\qquad S\,=\,0,\qquad \bar{B}_{1}\,+\,B_{2}\,=\,0, \qquad \bar{S}\,=\,0.
\end{eqnarray}
If we choose
$B_{2}\,=\,-\bar{B}_{1}\,=\,B$, we obtain the following expansions
\begin{eqnarray}
\Lambda^{(h)}(t, \theta , \bar{\theta})\,& = &\,\lambda (t)\,+\,\theta\,\dot{\bar{(C)}}\,
+\,\bar{\theta}\,(\dot{C})+\,\theta\,\bar{\theta}\,(i\,\dot{B})\nonumber\\
&\equiv &\lambda (t)\,+\,\theta\,(s_{ab}\,\lambda)\,+\,\bar{\theta}\,(s_{b}\,\lambda)\,
+\,\theta\,\bar{\theta}\,(s_{b}\,s_{ab}\,\lambda),\nonumber\\
F^{(h)}(t, \theta , \bar{\theta})\,& = &\,C(t) \,+\,\theta\,(-\,i\,B)\,+\,\bar{\theta}\,(0)
+\,\theta\,\bar{\theta}\,(0)\nonumber\\
&\equiv & C (t)\,+\,\theta\,(s_{ab}\,C)\,+\,\bar{\theta}\,(s_{b}\,C)\,
+\,\theta\,\bar{\theta}\,(s_{b}\,s_{ab}\,C),\nonumber\\
\bar{F}^{(h)}(t, \theta , \bar{\theta})\,& = &\,\bar{C} (t)\,+\,\theta\,(0)\,+
\,\bar{\theta}\,(i\,B)+\,\theta\,\bar{\theta}\,(0)\nonumber\\
&\equiv & \bar{C} (t)\,+\,\theta\,(s_{ab}\,\bar{C})\,+\,\bar{\theta}\,(s_{b}\,\bar{C})\,
+\,\theta\,\bar{\theta}\,(s_{b}\,s_{ab}\,\bar{C}),
\end{eqnarray}
where the superscript ($h$) denotes the expansions of the supervariables after the application
of the HC. The  above expansions of the supervariables lead to the derivation of proper 
(i.e. offshell nilpotent and absolutely anticommuting) (anti-)BRST symmetries that 
have been mentioned in (2). It is to be emphasized that the relationship $\bar B_1 + B_2 = 0$
in Eq. (13) is nothing but the CF-condition which turns out to be {\it trivial} for our Abelian theory. 
Furthermore, these expansions imply a relationship between 
the (anti-)BRST symmetries $s_{(a)b}$ and the translational generators 
($\partial_{\theta},\,\partial_{\bar\theta}$) along the Grassmannian directions of 
the (1, 2)-dimensional supermanifold. This relationship is:
$ lim_{\bar{\theta}\,=\,0}\,\partial_{\theta} \, \leftrightarrow \,\, 
s_{ab},  lim_{{\theta}\,=\,0}\,\partial_{\bar{\theta}}\leftrightarrow \,\, s_{b}$.
In view of this mapping, it is clear that we have the following equalities, namely;
\begin{eqnarray}
&&r(t)\,\rightarrow\, R(t, \theta, \bar{\theta})\,=\,r(t),\quad \quad B(t)\,
\rightarrow \mathcal{B}(t, \theta, \bar{\theta})\,=\,B(t),\nonumber\\
&&\theta(t)\,\rightarrow\,{\Theta}(t, \theta, \bar{\theta})\,=\,\theta(t),\quad 
\quad p_{\vartheta}(t)\,\rightarrow\,P_{\vartheta}(t, \theta, \bar{\theta})\,=\,p_{\vartheta}(t),
\end{eqnarray}
because of the fact that we have the (anti-)BRST invariance: $s_{(a)b}\,[r, \vartheta, p_{\vartheta}, B]\,=\,0$.

We now focus on the GIR that leads to the expansions of $P_{r}(t, \theta, \bar{\theta})$ 
(in terms of the basic and auxiliary variables) which is the generalizations of $p_{r}(t)$ 
onto the (1, 2)-dimensional supermanifold. We observe that $s_{(a)b}\,[\lambda\,+\,\dot{p}_{r}]\,=\,0$. 
It should be pointed out that all the gauge [i.e. (anti-)BRST]
invariant quantities are required to be independent of the Grassmannian variables when they are
generalized onto the appropriately chosen supermanifold.
This statement implies the following equality under the basic tenets of the augmented version of superfield
formalism (see, e.g., [11-14])
\begin{eqnarray}
\tilde{\lambda}^{(1)}_{(h)}\,+\,\tilde{d}\,P_{r}(t, \theta, \bar{\theta})\,=
\,\lambda^{(1)}(t)\,+\,dp_{r}(t),\qquad d\,=\,dt\,\partial_{t},
\end{eqnarray}
where ${\tilde{\lambda}^{(1)}}_{(h)}(t, \theta, \bar{\theta})\,=
\,dt\,\Lambda^{(h)}(t, \theta, \bar{\theta})\,+\,d\theta\,\bar{F}^{(h)}(t, \theta, \bar{\theta})\,
+\,d\bar{\theta}\,F^{(h)}(t, \theta, \bar{\theta})$. The super expansions for  
$\Lambda^{(h)}(t, \theta, \bar{\theta}), \bar{F}^{(h)}(t, \theta, \bar{\theta})$
and $F^{(h)}(t, \theta, \bar{\theta})$ are given in (14). Equating the coefficients 
of $dt, \, d\theta$ and $d\bar{\theta}$ from the l.h.s and r.h.s of Eq. (16), we obtain the 
following:
\begin{eqnarray}
&&\partial_{\theta}\,P_{r}(t, \theta, \bar{\theta})\,=\,-\,\bar C(t)\,-
\,i\,\bar{\theta}\,B(t)\,\equiv\,-\bar{F}^{(h)}(t, \theta, \bar{\theta}),\nonumber\\
&&\partial_{\bar{\theta}}\,P_{r}(t, \theta, \bar{\theta})\,=\,-\, C(t)\,+
\,i\,{\theta}\,B(t)\,\equiv\,-\,{F}^{(h)}(t, \theta, \bar{\theta}),\nonumber\\
&&\dot{P}_{r}(t, \theta, \bar{\theta})\,=\,\dot{p}_{r}(t)\,+\,\theta(-\,\dot{\bar{C}})\,+
\,\bar{\theta}(-\,\dot{C})\,+\,\theta\,\bar{\theta}(-\, i\,\dot{B}).
\end{eqnarray}
It is clear that the solution for the above conditions is the following relationship: 
\begin{eqnarray}
{P}_{r}(t, \theta, \bar{\theta})\,& = &\,p_r (t)\,+\,\theta(-\,\bar{C})\,+
\,\bar{\theta}(-\,{C})\,+\,\theta\,\bar{\theta}(-\,i\,{B})\nonumber\\
&\equiv & p_r (t)\,+\,\theta\,(s_{ab}\,p_{r})\,+\,\bar{\theta}\,(s_{b}\,p_{r})\,
+\,\theta\,\bar{\theta}\,(s_{b}s_{ab}\,p_{r}).
\end{eqnarray}
A careful  observation of the  Eqs. (14), (15) and (18) demonstrates that we have already 
derived the (anti-)BRST symmetry transformations (2) for the 1$D$ toy model of a rigid rotor.
We would like to emphasize that our present method of derivation is totally different from
our earlier works [17-19] on the supervariable approach to a 1$D$ rigid rotor.

We exploit the mathematical power of the HC in the context of the 2$D$ Abelian 1-form 
($A^{(1)}\,=\,dx^{\mu}\,A_{\mu}$) gauge theories of the modified versions of the 
Proca and anomalous gauge theories. 
In both these theories, we have the basic fields $A_{\mu}(x), C(x),$ and $\bar{C}(x)$ 
which can be generalized onto an appropriately chosen (2, 2)-dimensional supermanifold, parametrized by 
$ Z^{M}\,=\,(x^{\mu}, \theta, \bar{\theta})$. The generalizations of the gauge field $A_{\mu}(x)$ is
\begin{eqnarray}
A_{\mu}(x)\,\rightarrow\, B_{\mu}(x, \theta, \bar{\theta})\,=\,A_{\mu}(x)\,+
\,\theta\,\bar{R}_{\mu}(x)\,+\,\bar{\theta}\,R_{\mu}\,+\,i\,\theta\,\bar{\theta}\,S_{\mu},
\end{eqnarray}
and the generalizations of the fields: $C(x)\rightarrow\,F(x, \theta, \bar{\theta}),\,\, 
\bar{C}\rightarrow\,F(x, \theta, \bar{\theta})$ are given in Eq. (10) where we 
have to replace the parameter $t$ by the 2$D$ spacetime coordinate $x^{\mu}$ 
(with\,$\mu\,=\,0, 1$). Furthermore, we have the following generalizations
\begin{eqnarray}
&&d\,=\,dx^{\mu}\,\partial_{\mu}\rightarrow\,\tilde{d}\,=\,dx^{\mu}\,\partial_{\mu}\,+
\,d\theta\,\partial_{\theta}\,+\,d\bar{\theta}\,\partial_{\bar{\theta}},\nonumber\\
&&A^{(1)}\,=\,dx^{\mu}\,A_{\mu}\,\rightarrow\,\tilde{A}^{(1)}(x, \theta, \bar{\theta})\,
=\,dx^{\mu}\,B_{\mu}(x,\, \theta, \bar{\theta})\,+\,d\theta\,\bar{F}(x, \theta, \bar{\theta})\,
+\,d\bar{\theta}\,F(x, \theta, \bar{\theta}),
\end{eqnarray}
for the exterior derivative to super exterior derivative  and the connection 1-form 
to super connection 1-form on the (2, 2)-dimensional supermanifold.

The requirement of HC (i.e.\,$ \tilde{d}\,\tilde{A}^{(1)}\,=\,d\,A^{(1)}$) yields the
 following relationship between the secondary fields and basic and auxiliary fields 
of the Lagrangian densities (3) and (7)  for the Abelian 1-form gauge theories
under considerations, namely;
\begin{eqnarray}
&&R_{\mu}\,=\,\partial_{\mu}C, \qquad \quad \bar{R}_{\mu}\,=\,\partial_{\mu}\,\bar{C}, 
\qquad \quad S_{\mu}\,=\,\partial_{\mu}\,B, \nonumber\\
&&\bar{B}_{1}\,+\,B_{2}\,=\,0,\qquad \bar{B}_{2}\,=\,B_{1}\,=\,0, 
\qquad \bar{B}_{1}\,=\,B\,=\,-\,B_{2}.
\end{eqnarray}
We observe that $\bar B_1 + B_2 = 0$ is the {\it trivial}
(anti)-BRST invariant CF-condition which emerges out from our superfield formalism. This is 
one of the decisive features of our superfield approach.
The substitution of these values into the appropriate super expansions yields 
the following equations (see, e.g., [6-8,11-14] for details)
\begin{eqnarray}
B_{\mu}^{(h)}(x, \theta , \bar{\theta})\,& = &\,A_{\mu}(x)\,+
\,\theta\,(\partial_{\mu}\bar{C})\,+\,\bar{\theta}\,(\partial_{\mu}{C})\,
+\,\theta\,\bar{\theta}\,(i\,\partial_{\mu}\,B)\nonumber\\
&\equiv & A_{\mu}\,+\,\theta\,(s_{ab}\,A_{\mu})\,+\,\bar{\theta}\,(s_{b}\,A_{\mu})\,
+\,\theta\,\bar{\theta}\,(s_{b}\,s_{ab}\,A_{\mu}),\nonumber\\
F^{(h)}(x, \theta , \bar{\theta})\,& = &\,C(x)\,+\,\theta\,(-\,i\,B)\,+\,\bar{\theta}\,(0) \, 
+\,\theta\,\bar{\theta}\,(0)\nonumber\\
&\equiv &C(x)\,+\,\theta\,(s_{ab}\,C)\,+\,\bar{\theta}\,(s_{b}\,C)\,
+\,\theta\,\bar{\theta}\,(s_{b}\,s_{ab}\,C),\nonumber\\
\bar{F}^{(h)}(x, \theta , \bar{\theta})\,& = &\,\bar{C}(x)\,+\,\theta\,(0)\,+
\,\bar{\theta}\,(i\,B)+\,\theta\,\bar{\theta}\,(0)\nonumber\\
&\equiv &\bar{C}(t)\,+\,\theta\,(s_{ab}\,\bar{C})\,+\,\bar{\theta}\,(s_{b}\,\bar{C})\,
+\,\theta\,\bar{\theta}\,(s_{b}\,s_{ab}\,\bar{C}),
\end{eqnarray}
where the superscript ($h$) denotes the expansions of the superfields after
 the application of the HC. The expansions (22) are common to all the 2$D$ theories 
of our interest. It is clear that we have the mappings: 
$s_{b}\,\Leftrightarrow \,\partial_{\bar{\theta}}, \, s_{ab}\,\Leftrightarrow \,\partial_{\theta}$ 
as pointed out earlier.

We now focus on the derivation of the (anti-)BRST symmetries for the Stueckelberg field 
$\phi(x, t)$ of the modified version of the 2$D$ Proca theory. We observe that 
$s_{(a)b}\,[A_{\mu}\,-\,\frac{1}{m}\,\partial_{\mu}\,\phi]\,=\,0$.
Thus, we have the following restrictions due to the basic tenets of the
augmented version of superfield formalism (see, e.g., [20,21] for details)
\begin{eqnarray}
\tilde{A}^{(1)}_{(h)}(x, \theta, \bar{\theta})\,-\,\frac{1}{m}\,\tilde{d}\,
\Phi(x, \theta, \bar{\theta})\,=\,A^{(1)}(x)\,-\,\frac{1}{m}\,d\,\phi(x), 
\qquad \quad d\,=\,dx^{\mu}\,\partial_{\mu}
\end{eqnarray}
where $\tilde{A}^{(1)}_{(h)}\,=\,dx^{\mu}\,B_{\mu}^{(h)}(x, \theta, \bar{\theta})\,
+\,d\theta\,\bar{F}^{(h)}(x, \theta, \bar{\theta})\,+\,d\bar{\theta}\,{F}^{(h)}(x, \theta, \bar{\theta})$ 
and $A^{(1)}\,=\,dx^{\mu}\,A_{\mu}(x)$. Exploiting the expansions of (22), we obtain the following conditions
\begin{eqnarray}
&&\partial_{\theta}\,\Phi(x, \theta, \bar{\theta})\,= \, m\,\bar{F}^{(h)}(x, \theta, \bar{\theta})\,
\equiv\, m \, \bar{C}(x)\,+\,\bar{\theta}\,(i\, m \,B),\nonumber\\
&&\partial_{\bar{\theta}}\,\Phi(x, \theta, \bar{\theta})\,=\, m \,{F}^{(h)}(x, \theta, \bar{\theta})\,
\equiv\, m \,{C}(x)\,+\,{\theta}\,(-\,i\, m \,B),\nonumber\\
&&\partial_{\mu}\,\Phi(x, \theta, \bar{\theta})\,=\,\partial_{\mu}\,\phi(x)\,+
\,\theta\,( m\,\partial_{\mu}\,\bar{C})\,+\,\bar{\theta}\,(m\,\partial_{\mu}\,C)\,+
\,\theta\,\theta\,(i\,m\,\partial_{\mu}\,B),
\end{eqnarray}
from the equation (23). The solution of the above conditions is the following
\begin{eqnarray}
\Phi^{(g)}\,(x, \theta, \bar{\theta})\,& = &\,\phi(x)\,+\,\theta\,(m\,\bar{C})\,
+\,\bar{\theta}\,(m\,C)\,+\,\theta\,\bar{\theta}\,(i\,m\,B)\nonumber\\
&\equiv & \phi\,(x)\,+\,\theta\,(s_{ab}\,\phi)\,+\,\bar{\theta}\,(s_{b}\,\phi)\,
+ \,\theta\,\bar{\theta}\,(s_{b}\,s_{ab}\phi),
\end{eqnarray}
where the superscript ($g$) denotes the expansions of the above superfields after the 
application of the GIR in (23). Thus, it is evident that we have already obtained 
the (anti-)BRST symmetry transformations for the field $\phi(x)$. The (anti-)BRST 
invariance $s_{(a)b}\,B\,=\,0$ of the Nakanishi-Lautrup type field $B(x)$ 
implies that we have: $B(x)\,\rightarrow\,\mathcal{B}(x, \theta, \bar{\theta})\,=\,B(x)$.

A close look at the (anti-)BRST symmetry transformations (6) shows that 
$s_{(a)b}\,[\lambda\,+\,\dot{\phi}]\,=\,0$ and  $s_{(a)b}\,[\lambda\,+\,\dot{v}]\,=\,0$ 
in the case of the BRST description of the 2$D$ self-dual bosonic theory. As a consequence,
we have the following equality due to the basic tenets of the augmented version of 
superfield formalism (see, e.g., [22,23] for details)
\begin{eqnarray}
&&\tilde \lambda^{(1)}_{(h)}\,(x, \theta, \bar{\theta})\,+\,\tilde{d}\,
\Phi(x, \theta, \bar{\theta})\,=\,\lambda^{(1)}(x)\,+\,d\,\phi(x),\nonumber\\
&&\tilde \lambda^{(1)}_{(h)}\,(x, \theta, \bar{\theta})\,+\,\tilde{d}\,V(x, \theta, \bar{\theta})\,
=\,\lambda^{(1)}(x)\,+\,d\,v(x),
\end{eqnarray}
where $d\,=\,dt\,\partial_{t}$ and $\tilde{d}\,=\,dt\,\partial_{t}\,+\,d\theta\,\partial_{\theta}\,
+\,d\bar{\theta}\,\partial_{\bar{\theta}}$. It will be noted that, even though our theory is a two 
(1 + 1)-dimensional theory, it is the coordinate $t$ that plays the role of the evolution parameter [23]. 
We emphasize  that $\tilde{\lambda}^{(1)}_{(h)}\,=\,dt\,\lambda^{(h)}(x, \theta, \bar{\theta})\,
+\,d\theta\,\bar{F}^{(h)}(x, \theta, \bar{\theta})\,+\,d\bar{\theta}\,{F}^{(h)}(x, \theta, \bar{\theta})$
 is the same as in the case of a rigid rotor with the replacement $t\,\rightarrow\,x^{\mu}$. 
It is worthwhile to point out that the 2$D$ self-dual chiral bosonic
theory is different from the 1$D$ toy model of a rigid rotor because here the basic 
fields $\lambda, \, C, \, \bar C$ are functions of the 2$D$ spacetime coordinates ($x^\mu$).
The equality (26) leads to the following relationships:
\begin{eqnarray}
&&\partial_{\theta}\,V(x, \theta, \bar{\theta})\,\equiv\,\partial_{\theta}\,\Phi(x,\theta, \bar{\theta})\,
=\,-\,\bar{F}^{(h)}(x, \theta, \bar{\theta})\,=\,-\,\bar{C}(x)\,-\,i\,\bar\theta\,B(x),\nonumber\\
&&\partial_{\bar{\theta}}\,V(x, \theta, \bar{\theta})\,\equiv\,\partial_{\bar{\theta}}\,
\Phi(x,\theta, \bar{\theta})\,=\,-\,{F}^{(h)}(x, \theta, \bar{\theta})\,=\,-\,{C}(x)\,+
\,i\,\theta\,B(x),\nonumber\\
&&\dot{\Phi}(x, \theta\, \bar{\theta})\,=\,\dot{\phi}(x)\,+\,\theta\,(-\,\dot{\bar{C}})\,+
\,\bar{\theta}\,(-\,\dot{C})\,+\,\theta\,\bar{\theta}\,(-\,i\,\dot{B}),\nonumber\\
&&\dot{V}(x, \theta\, \bar{\theta})\,=\,\dot{v}(x)\,+\,\theta\,(-\,\dot{\bar{C}})\,+
\,\bar{\theta}\,(-\,\dot{C})\,+\,\theta\,\bar{\theta}\,(-\,i\,\dot{B}).
\end{eqnarray}
The above relations imply that we have the following super expansions for the superfields 
$\Phi (x, \theta, \bar{\theta}) $ and $V (x, \theta, \bar{\theta})$ after the application 
of the GIRs (26), namely;
\begin{eqnarray}
\Phi^{(g)}(x, \theta, \bar{\theta})\,& = &\,\phi(x)\,+\,\theta\,(-\,\bar{C})\,+
\,\bar{\theta}\,(-\,C)\,+\,\theta\,\bar{\theta}\,(-\,i\,B)\nonumber\\
&\equiv & \phi(x)\,+\,\theta\,(s_{ab}\,\phi)\,+\,\bar{\theta}\,(s_{b}\,\phi)\,+
 \,\theta\,\bar{\theta}\,(s_{b}\,s_{ab}\phi),\nonumber\\
V^{(g)}(x, \theta, \bar{\theta})\,& = &\,v(x)\,+\,\theta\,(-\,\bar{C})\,+
\,\bar{\theta}\,(-\,C)\,+\,\theta\,\bar{\theta}\,(-\,i\,B)\nonumber\\
&\equiv & v(x)\,+\,\theta\,(s_{ab}\,v)\,+\,\bar{\theta}\,(s_{b}\,v)\,+
\,\theta\,\bar{\theta}\,(s_{b}\,s_{ab}\,v).
\end{eqnarray}
Thus, we have determined all the (anti-)BRST symmetry transformations for the self-dual 
2$D$ bosonic theory as is evident from equations (14) and (28).

Finally, we note that the (anti-)BRST invariant combinations
 $s_{(a)b}\,[A_{\mu}\,+\,\partial_{\mu}\,\phi]\,=\,0$ and 
$s_{(a)b}\,[A_{\mu}\,-\,\partial_{\mu}\,\sigma]\,=\,0$, in some sense, 
are the physical quantities in the 2$D$ modified version of anomalous gauge theory.
As a consequence, we have the following equalities due to the basic concepts of the augmented 
version of superfield approach (see, e.g. [11-14] for details)
\begin{eqnarray}
&&\tilde A^{(1)}_{(h)}\,(x, \theta, \bar{\theta})\,+\,\tilde{d}\,
\Phi\,(x, \theta, \bar{\theta})\,=\,A^{(1)}(x)\,+\,d\,\phi(x),\nonumber\\
&&\tilde A^{(1)}_{(h)}\,(x, \theta, \bar{\theta})\,-\,\tilde{d}\,\Sigma 
\,(x, \theta, \bar{\theta})\,=\,A^{(1)}(x)\,-\,d\,\sigma(x),
\end{eqnarray}
where the 2$D$ fields $\phi(x)$ and $\sigma(x)$ have been generalized to the superfields on the 
(2, 2)-dimensional supermanifold as: $\phi(x) \rightarrow \Phi(x, \theta, \bar{\theta}),
 \, \sigma(x) \to \Sigma (x, \theta, \bar{\theta})$ and
$\tilde{A}^{(1)}_{(h)} = dx^{\mu}\,B_{\mu}^{(h)} (x, \theta, \bar{\theta})
+ d{\theta}\,\bar{F}^{(h)}\,(x, \theta, \bar{\theta}) + d\bar{\theta}\,{F}^{(h)}\,
(x, \theta, \bar{\theta})$. The equalities in (29) lead to the following equations:
\begin{eqnarray}
&&\partial_{\theta}\,\Sigma\,(x, \theta, \bar{\theta})\,\equiv\,-\,\partial_{\theta}\,
\Phi\,(x, \theta, \bar{\theta})\,=\,\bar{F}^{(h)}\,(x, \theta, \bar{\theta}), \nonumber\\
&&\partial_{\bar{\theta}}\,\Sigma\,(x, \theta, \bar{\theta})\,\equiv\,-\,
\partial_{\bar{\theta}}\,\Phi\,(x, \theta, \bar{\theta})\,=\,{F}^{(h)}\,
(x, \theta, \bar{\theta}),\nonumber\\
&&\partial_{\mu}\,\Sigma\,(x, \theta, \bar{\theta})\,=\,\partial_{\mu}\,
\sigma\,+\,\theta\,(\partial_{\mu}\,\bar{C})\,+\,\bar{\theta}\,(\partial_{\mu}\,C)\,
+\,\theta\,\bar{\theta}\,(i\,\partial_{\mu}\,B),\nonumber\\
&&\partial_{\mu}\,\Phi\,(x, \theta, \bar{\theta})\,=\,\partial_{\mu}\,\phi\,+
\,\theta\,(-\,\partial_{\mu}\,\bar{C})\,+\,\bar{\theta}\,(-\, \partial_{\mu}\,C)\,
+\,\theta\,\bar{\theta}\,(-\,i\,\partial_{\mu}\,B).
\end{eqnarray}
The above conditions are satisfied by the following explicit super expansions for the super fields
(corresponding to the ordinary fields $\phi (x)$ and $\sigma (x)$), namely;
\begin{eqnarray}
\Phi^{(g)}(x, \theta, \bar{\theta})\,& = &\,\phi(x)\,+\,\theta\,(-\,\bar{C})\,+
\,\bar{\theta}\,(-\,C)\,+\,\theta\,\bar{\theta}\,(-\,i\,B)\nonumber\\
&\equiv & \phi(x)\,+\,\theta\,(s_{ab}\,\phi)\,+\,\bar{\theta}\,(s_{b}\,\phi)\,+
 \,\theta\,\bar{\theta}\,(s_{b}\,s_{ab}\phi),\nonumber\\
\Sigma^{(g)}(x, \theta, \bar{\theta})\,& = &\,\sigma(x)\,+\,\theta\,(\bar{C})\,+
\,\bar{\theta}\,(C)\,+\,\theta\,\bar{\theta}\,(i\,B)\nonumber\\
&\equiv & \sigma (x)\,+\,\theta\,(s_{ab}\,\sigma (x))\,+\,\bar{\theta}\,(s_{b}\,\sigma (x))\,
+ \,\theta\,\bar{\theta}\,(s_{b}\,s_{ab}\,\sigma (x)).
\end{eqnarray}
where the superscript ($g$) denotes the super expansions obtained after the application of 
the GIRs mentioned in (29). Finally, we know that $s_{(a)b}\,B(x)\,=\,0$. As a consequence,
 it is evident that we have the generalization: 
$B(x)\,\rightarrow \,\mathcal{B}(x, \theta, \bar{\theta})\,=\,B(x)$. 
A close and careful look at (22) and (31) demonstrates that we have obtained {\it all} 
the (anti-)BRST  transformations for {\it all} the fields of the 2$D$ modified 
version of the anomalous gauge theory.

\section{SUSP Unitary Operator: Universality Aspects}

We establish, in this section, that the form of the SUSP unitarity operator 
(and its hermitian conjugate) is the {\it same} for {\it all} the Abelian models 
we have discussed in our present investigation and the 4$D$ interacting  
$U(1)$ gauge theory with Dirac and complex scalar fields that have been discussed 
in our earlier work [15] where we have derived the explicit form of these operators
 (for the electric charge $e\,=\,1$) as (see, e.g., [15]) for details):
\begin{eqnarray}
&&U\,(x, \theta, \bar{\theta})\,=\,1+\,\theta\,(-\,i\,\bar{C})\,+\,\bar{\theta}\,(-\,i\,C)\,
+\,\theta\,\bar{\theta}\,(B\,-\,C\,\bar{C}),\nonumber\\
&&U^{\dagger}\,(x, \theta, \bar{\theta})\,=\,1+\,\theta\,(i\,\bar{C})\,+\,\bar{\theta}\,(i\,C)\,
+\,\theta\,\bar{\theta}\,(-\,B\,+\,\bar{C}\,{C}).
\end{eqnarray}
The above unitarity operators can be exponentiated as follows [15]
\begin{eqnarray}
&&U\,(x, \theta, \bar{\theta})\,=\,exp\,[\theta\,(-\,i\,\bar{C})\,+
\,\bar{\theta}\,(-\,i\,C)\,+\,\theta\,\bar{\theta}\,B],\nonumber\\
&&U^{\dagger}\,(x, \theta, \bar{\theta})\,=\,exp\,[\theta\,(i\,\bar{C})\,+
\,\bar{\theta}\,(i\,C)\,-\,\theta\,\bar{\theta}\,B].
\end{eqnarray} 
These forms of the unitary operators explicitly show the Abelian $U(1)$ group structure in
the transformation superspace of our theory and it is elementary to check that 
$U\,U^{\dagger}\,=\,U^{\dagger}\,U\,=\,1$. The latter can be also checked from 
the explicit expansions of $U(x, \theta, \bar{\theta})$ and 
$U^{\dagger}(x, \theta, \bar{\theta})$ which are quoted in (32). In fact,
the group structure becomes clear due to (33).

In the case of 1$D$ toy model of a rigid rotor, it is very interesting to note that the HC
can be expressed as follows
\begin{eqnarray}
\tilde{\lambda}^{(1)}_{(h)}\,=\,U\,(t, \theta, \bar{\theta})\,\lambda^{(1)}\,(t)\,
U^{\dagger}\,(t, \theta, \bar{\theta})\, +\,i\,\tilde{d}\,U\,(t, \theta, \bar{\theta})
\,U^{\dagger}\,(t, \theta, \bar{\theta}),
\end{eqnarray} 
where the expression for $U(t, \theta, \bar{\theta})$ is same as given in (32) and (33) 
with the replacement: $U\,(x, \theta, \bar{\theta }){\mid} _{x\,=\,t}\,\equiv \, 
U(t, \theta, \bar{\theta})$ and the other symbols in (34) have already been expalined earlier.
It is worthwhile to mention that the relationship (34) is just the 1$D$ version of 
{\it exactly} the same kind of relationship (cf. (40) below) obtained in our earlier
work on the interacting 4$D$ Abelian $U(1)$ gauge theory with Dirac and complex scalar fields [15].
We can  readily check that the equation (34) provides an alternative to the HC 
($\tilde{d}\,\tilde{\lambda}^{(1)}_{(h)}\,=\,d\,\lambda^{(1)}$)
because we get the exact form of the super expansions (14) which have already been obtained 
earlier in Sec. 3. Furthermore, our Eq. (34) {\it also} provides the reason behind 
the application of the HC because it can be explicitly checked that
\begin{eqnarray}
\tilde{d}\,\tilde{\lambda}^{(1)}_{(h)}\,=\,\tilde{d}\,\lambda^{(1)}\,(t)\,
-\,i\,\tilde{d}\,U\,(t, \theta, \bar{\theta})\,\wedge \tilde{d}\,U^{\dagger}\,(t, \theta, \bar{\theta}),
\end{eqnarray} 
where we have applied the operator $\tilde{d}\,=\,dt\,\partial_{t}\,+\,d\theta\,\partial_{\theta}\,+
\,d\bar{\theta}\,\partial_{\bar{\theta}}$ from the left on equation (34) and used the 
basic property of the differential geometry [(i.e. $\tilde{(d)}^{2}\,=\,0$)]. We note that 
$\tilde{d}\,\tilde{\lambda}^{(1)}\,(t)\,=\,\tilde{d}\,\lambda^{(1)}\,(t)$ 
(as $ \partial_{\theta}\,\lambda^{(1)}\,(t)\,=\,\partial_{\bar{\theta}}\,\lambda^{(1)}\,(t)\,=\,0$)
and it is  the forms of $U$ and $U^{\dagger}$ in (33) that shows that 
($\tilde d\,U\,\wedge \,\tilde d\,{U}^{\dagger}\,=\,0$). This statement can  be also verified by 
the explicit expansions as listed below:
\begin{eqnarray}
\tilde d\,U &=& dt\,\bigl[\theta (-\, i\, \dot{\bar C}) + \bar\theta\, (-\, i\, \dot{C}) + 
\theta\,\bar\theta (\dot B - \dot C\, \bar C - C\,\dot{\bar C})\bigr] \nonumber\\ &+& 
d\theta \bigl[ - \,i\,\bar C + \bar\theta (B - C\, \bar C) \bigr] +
d \bar\theta \bigl[ - \,i\, C - \theta (B - C\, \bar C) \bigr], \nonumber\\
\tilde d\,U^\dagger &=& dt\,\bigl[\theta (i\, \dot{\bar C}) + \bar\theta\, (i\, \dot{C}) + 
\theta\,\bar\theta (-\, \dot B + \dot {\bar C}\, C + \bar C\,\dot{C})\bigr] \nonumber\\ &+& 
d\theta \bigl[i\,\bar C + \bar\theta (-\, B + \bar C\, C) \bigr] +
d \bar\theta \bigl[i\, C - \theta (-\,B + \bar C\, C) \bigr]. 
\end{eqnarray}
We can collect the coefficients of $(dt \wedge d\theta), \, (d\theta \wedge d\bar\theta), \,
(d\theta \wedge d\theta), \, (d\theta \wedge d\bar\theta)$ and $(d\bar\theta \wedge d\bar\theta)$ in
$(\tilde d\,U \wedge \tilde d\,U^\dagger)$. Interestingly, these coefficients turn out to be 
{\it exactly} zero. This observation, once again, proves that $\tilde d\,U\, \wedge \,\tilde d\,U^\dagger = 0$
where, of course, we have used the input $dt \wedge dt = 0$.

The GIRs (i.e. (anti-)BRST) restrictions $s_{(a)b} \bigl[\lambda + {\dot p}_r \bigr] = 0$ 
have been exploited to derive the expansion for $P^{(g)}_r (t, \theta, \bar{\theta})$ in Eq. (18). 
This relationship can be also expressed in terms of the unitary operators (with $p_r (t) \to 
P_r (t, \theta, \bar{\theta})$) because the equality
\begin{eqnarray}
\tilde d\, P_r(t, \theta, \bar{\theta}) + {\tilde \lambda}^{(1)}_{(h)} (t, \theta, \bar{\theta})
 = d \, p_r (t) + \lambda^{(1)} (t),
\end{eqnarray}
leads to the following relationships when we use (34) in it, namely;
\begin{eqnarray}
&&\partial_\theta P_r (t, \theta, \bar{\theta}) = -\,i\, (\partial_\theta \, U) \, U^\dagger \equiv -\, 
\bar F^{(h)} (t, \theta, \bar{\theta}),\nonumber\\
&&\partial_{\bar\theta} P_r (t, \theta, \bar{\theta}) = -\,i\, (\partial_{\bar\theta} \, U) \, U^\dagger \equiv -\, 
F^{(h)} (t, \theta, \bar{\theta}),\nonumber\\
&& \partial_t \, P_r (t, \theta, \bar{\theta}) = \partial_t p_r  -\,i\, (\partial_t \, U) \, U^\dagger .
\end{eqnarray}
The last relationship yields the following explicit equation
\begin{eqnarray}
{\dot P}_r (t, \theta, \bar{\theta}) = {\dot p}_r  (t) + \theta \, (-\, \dot{\bar C}) + \bar\theta \, (-\, \dot{C}) 
+ \theta\, \bar\theta (-\,i\, \dot B). 
\end{eqnarray}
The solution of (38) and (39) is nothing but he expansion given in (18) which leads to the
derivation of the (anti-)BRST symmetry transformations for $p_r (t)$.

For the modified versions of the 2$D$ Proca and anomalous Abelian 1-form gauge theories, we have the
following form of the HC in terms of the SUSP unitary operators, namely;
\begin{eqnarray}
{\tilde A}_{(h)}^{(1)} (x, \theta, \bar{\theta}) = U (x, \theta, \bar{\theta}) \, A^{(1)} (x)
 \,U^\dagger (x, \theta, \bar{\theta}) 
+ i\, ({\tilde d}\, U (x, \theta, \bar{\theta}))\, U^\dagger (x, \theta, \bar{\theta}),
\end{eqnarray}
where the forms of $U (x, \theta, \bar{\theta})$ and $U^\dagger (x, \theta, \bar{\theta}) $
are quoted in (32) and (33). We wish to lay emphasis on the fact that the above relationship has
been obtained in the 4$D$ {\it interacting} Abelian 1-form $U(1)$ gauge theory with Dirac and complex scalar
fields where the concept of the covariant derivatives plays an important role. The point to be noted 
(and emphasized) is the observation that the relationship (40) is valid for {\it all} kinds of 
Abelian 1-form gauge theory where even the  {\it covariant derivatives} are {\it not} explicitly defined.
The substitution of the explicit forms of SUSP unitary operators from (32) and (33) into (40) produces 
the expansions that are given in (22) where one has to take into account the definition of the 
super 1-form: ${\tilde A}^{(1)}_{(h)} = dx^\mu \, B^{(h)}_\mu (x, \theta, \bar{\theta}) +
 d\theta \, {\bar F}^{(h)} (x, \theta, \bar{\theta}) + d\bar\theta \,
{F}^{(h)} (x, \theta, \bar{\theta})$ on the l.h.s.

The existence of SUSP unitary operators and its hermitian conjugate also provides the logical
reason behind the imposition of the HC (cf. Sec. 3) as it can be shown that
\begin{eqnarray}
\tilde d \,{\tilde A}^{(1)}_{(h)} = \tilde d \, {A}^{(1)} (x) - i\, (\tilde d\, U) \wedge (\tilde d \, U^\dagger).
\end{eqnarray}
The above equation emerges out from (40) when we apply a super exterior derivative $\tilde d$ from 
the left on it. It is evident that $\tilde d \, {A}^{(1)} (x) =  d \, {A}^{(1)} (x)$ and 
$(\tilde d\, U) \wedge (\tilde d \, U^\dagger) = 0$ due to the form of $U (x, \theta, \bar{\theta})$
and $U^\dagger (x, \theta, \bar{\theta})$ in (33). One can explicitly compute the expression
$(\tilde d\, U) \wedge (\tilde d \, U^\dagger)$ and collect the coefficients of $(dx^\mu \wedge dx^\nu), \,
(dx^\mu \wedge d\theta), \, (dx^\mu \wedge d\bar\theta), \, (d\theta \wedge d\theta),\,
(d\bar\theta \wedge d\bar\theta),\, (d\theta \wedge d\bar\theta)$ to prove that all these coeficients are zero. 
Thus, we finally obtain the HC as: $\tilde d \,{\tilde A}^{(1)}_{(h)} =  d \, {A}^{(1)} (x)$ which
primarily emerges out from the equation (40) that is expressed in terms of the SUSP unitary 
operators $U (x, \theta, \bar{\theta})$ and $U^\dagger (x, \theta, \bar{\theta})$. The key equations (40), (41)
and the HC ($\tilde d \, {\tilde A}^{(1)} =  d \, {A}^{(1)}$) are common features to the BRST description of the 2$D$
modified versions of Proca and anomalous gauge theories. On the other hand, the 1$D$ rigid rotor and 2$D$ self-dual
field theory are described by the relationships (34) and (35) 
(with $\tilde d \, {\tilde{\lambda}}^{(1)}_{(h)}  =  d \, {\lambda}^{(1)}$). However, in the case of the 2$D$
self-dual theory, one has to replace $t$ by $x^{\mu} \, (\mu = 0, 1)$ in the expression for the 
gauge variable $\lambda(t)$ (e.g. $\lambda (t) \, \to \lambda (x, t)$, etc.).

We are now in the position to express the GIR in the language of SUSP unitary operator 
and its hermitian conjugate. First of all, we focus on the modified version of 2$D$ Proca theory
and exploit the relationship given in (23). Using the relationship of Eq. (40), we obtain the
following 
\begin{eqnarray}
&&\partial_\theta \, \Phi (x, \theta, \bar{\theta}) = i\, m \, (\partial_\theta \, U)\, U^\dagger, \nonumber\\
&&\partial_{\bar\theta} \, \Phi (x, \theta, \bar{\theta}) = i\,  m \,(\partial_{\bar\theta} \, U)\, U^\dagger, \nonumber\\
&&\partial_\mu \, \Phi (x, \theta, \bar{\theta}) = \partial_\mu \, \phi (x) + i\, m \, (\partial_\mu \, U)\, U^\dagger,
\end{eqnarray}
in the language of the SUSP unitary operator and its hermitian conjugate. The substitution of 
the explicit form of $U (x, \theta, \bar{\theta})$ and $U^\dagger (x, \theta, \bar{\theta})$
from (32) leads to the solution of (42) as given in (25) (see, Sec. 3). As far as,
the GIRs (26) for the 2$D$ self-dual theory is concerned, we can express these in terms of the 
SUSP unitary operators as:
\begin{eqnarray}
&&\partial_\theta \, V (x, \theta, \bar{\theta}) \equiv \partial_\theta \, \Phi (x, \theta, \bar{\theta})
= -\,i\, (\partial_\theta \, U)\, U^\dagger, \nonumber\\
&&\partial_{\bar\theta} \, V (x, \theta, \bar{\theta}) \equiv \partial_{\bar\theta} \, \Phi (x, \theta, \bar{\theta})
= -\,i\, (\partial_{\bar\theta} \, U)\, U^\dagger, \nonumber\\
&&\partial_t \, V (x, \theta, \bar{\theta}) = \partial_t \, v (x) - i\, (\partial_t \, U)\, U^\dagger, \nonumber\\
&&\partial_t \, \Phi (x, \theta, \bar{\theta}) = \partial_t \, \phi (x) - i\, (\partial_t \, U)\, U^\dagger,
\end{eqnarray}
where we have used the definition: $\tilde{d}\,=\,dt\,\partial_{t}\,+\,d\theta\,\partial_{\theta}\,+
\,d\bar{\theta}\,\partial_{\bar{\theta}}$ in the equation (26) because the evolution parameter of this
2$D$ theory is $t$ only (see, e.g. [23], for details). The explicit substitution of 
$U (x, \theta, \bar{\theta})$ and $U^\dagger (x, \theta, \bar{\theta})$ into (43) yields the 
relation (27) whose solution is (28) in terms of the (anti-)BRST symmetry transformations.

Finally, we concentrate on the GIRs listed in (29) for the 2$D$ modified version of the 
anomalous gauge theory. Using the equation (40) in (29), we obtain the following:
\begin{eqnarray}
&&\partial_\theta \, \Sigma (x, \theta, \bar{\theta}) \equiv  - \, \partial_\theta \, \Phi (x, \theta, \bar{\theta})
= i\, (\partial_\theta \, U)\, U^\dagger, \nonumber\\
&&\partial_{\bar\theta} \, \Sigma (x, \theta, \bar{\theta}) \equiv -\, \partial_{\bar\theta} \, \Phi (x, \theta, \bar{\theta})
= i\, (\partial_{\bar\theta} \, U)\, U^\dagger, \nonumber\\
&&\partial_\mu\, \Sigma (x, \theta, \bar{\theta}) = \partial_\mu \, \sigma (x) + i\, (\partial_\mu \, U)\, U^\dagger, \nonumber\\
&&\partial_\mu \, \Phi (x, \theta, \bar{\theta}) = \partial_\mu \, \phi (x) - i\, (\partial_\mu \, U)\, U^\dagger.
\end{eqnarray}
It is straightforward to check that the r.h.s. of (44) yields the r.h.s. of (30) whose
solution is (31). Thus, we conclude that the GIRs, used in the cases of the 1$D$ and 2$D$ models under 
consideration, can {\it all} be expressed in terms of the SUSP unitary operator and its 
hermitian conjugate. As a consequence, these operators do provide the alternatives to 
the HC and the GIRs that are exploited in the derivation of the full set of 
(anti-)BRST symmetries for the models under consideration.

\section{Conclusions}

We have established, in our present investigation, that the SUSP unitary operator and its hermitian
conjugate have a {\it universal} mathematical expression which is true for different kinds of
$U(1)$ Abelian 1-form gauge theories. In the case of 4$D$ interacting Abelian $U(1)$ gauge theory with the
Dirac and complex scalar fields (where the covariant derivatives are explicitly defined), the SUSP
unitary operator and its hermitian conjugate have been derived explicitly within the framework of
augmented version of superfield formalism [15]. It turns out that {\it this} mathematical form of 
the above operators remains relevant and correct even in the context of the Abelian 1-form gauge
theories in the one (0 + 1)-dimension and two (1 + 1)-dimensions of spacetime, too, where the covariant 
derivatives are {\it not} defined in an explicit manner. The observation
of the {\it universality} of these operators, we re-emphasize, is one of the highlights of our present
investigation.

In our present endeavor, we have applied the augmented version of superfield formalism [15] 
to derive the proper (i.e. off-shell nilpotent and absolutely anticommuting) (anti-) BRST 
symmetry transformations for a {\it new} system of the 2$D$ modified version of anomalous gauge theory
(see, Sec. 3). Thus, it is a {\it novel} result in our present investigation. We have established that
our superfield approach could be applied to the new systems of field theories and/or the toy models of
gauge theories and we can be sure of obtaining the proper (anti-) BRST symmetry transformations.
The key ingredients that play a decisive role in these derivations are the celebrated HC and GIRs.
One of the key features of our superfield approach is the emergence of the CF-condition (cf. (13), (21)), which turns
out to be {\it trivial} in the case of Abelian 1-form gauge theories of different varieties (cf. (13), (21)).

The toy model and the field theoretical examples, considered in our present investigation,
are interesting and instructive because their {\it internal} symmetries provide the physical realizations of the
de Rham cohomological operators of differential geometry within the framework of the BRST formalism. For these
models of gauge theories [17-23], there exist {\it six} continuous symmetries and a couple of discrete symmetries
which provide the analogues of the cohomological operators, the Hodge duality operation, 
the properties of the differential operators, etc.,  in a physical manner (see, e.g., [17-23] for details).
Thus, these models have a whole lot of richness (as far as the mathematical and physical 
aspects of a well-defined physical theory in any arbitrary dimension of spacetime is concerned).

It would be a very nice future endeavor to establish the universality of the SUSP operators 
in the context of some models of the non-Abelian nature. The models, we have considered in our present 
investigation, are examples of  Hodge theory where the proper (anti-)co-BRST symmetries {\it also} 
exists. Thus, it would be a challenging problem for us to derive the SUSP  unitary operator
and its hermitian conjugate corresponding to these proper continuous symmetry transformations. We 
hope that the discrete symmetry transformations in our models for the Hodge theory (corresponding to the Hodge 
duality ($*$) operation) would be able to lead us to derive these SUSP unitary operators
 in a consistent and cogent manner. It would be 
also interesting to obtain such operators in the context of diffeomorphism  invariant theories. 
We are currently  busy with such kind of issues which, we hope, to resolve in our future publications.\\

\noindent
{\bf Acknowledgement:} One of us (TB) would like to gratefully acknowledge the financial 
support from CSIR, Govt of India, New Delhi, under its SRF-scheme. Another author (NS) is thankful to the 
BHU-fellowship for financial support. The present investigation has been carried out under
the above financial supports.\\

\end{document}